\documentclass[conference]{IEEEtran}
\IEEEoverridecommandlockouts

\usepackage{cite}
\usepackage{amsmath,amssymb,amsfonts}
\usepackage{graphicx}
\usepackage{textcomp}
\usepackage{xcolor}
\usepackage{amsmath}
\usepackage{caption}
\usepackage{subcaption}
\usepackage{graphicx}  
\usepackage{dcolumn}   
\usepackage{bm}        
\usepackage{relsize}  
\usepackage{verbatim} 
\usepackage{float}
\usepackage{mathtools}
\usepackage{color}
\usepackage{svg}
\usepackage{import}
\usepackage{comment}
\usepackage{tikz}
\usepackage{algorithm}
\usepackage{algpseudocode}

\newcommand{\review}[1]{#1}

\def\BibTeX{{\rm B\kern-.05em{\sc i\kern-.025em b}\kern-.08em
    T\kern-.1667em\lower.7ex\hbox{E}\kern-.125emX}}
\begin{document}

\title{Fourier neural operators for spatiotemporal dynamics in two-dimensional turbulence
}


\author{
\IEEEauthorblockN{
    Mohammad Atif\IEEEauthorrefmark{1},
    Pulkit Dubey\IEEEauthorrefmark{2},
    Pratik P. Aghor\IEEEauthorrefmark{3},
    Vanessa L{\'o}pez-Marrero\IEEEauthorrefmark{1},
    Tao Zhang\IEEEauthorrefmark{1},\\
    Abdullah Sharfuddin\IEEEauthorrefmark{4},
    Kwangmin Yu\IEEEauthorrefmark{1},
    Fan Yang\IEEEauthorrefmark{1},
    Foluso Ladeinde\IEEEauthorrefmark{4}, 
    Yangang Liu\IEEEauthorrefmark{1},
    Meifeng Lin\IEEEauthorrefmark{1},\\
    Lingda Li\IEEEauthorrefmark{1}
    }
       
    \IEEEauthorblockA{\IEEEauthorrefmark{1}Brookhaven National Laboratory, NY, USA}
    \IEEEauthorblockA{\IEEEauthorrefmark{2}University of New Hampshire, NH, USA} 
    \IEEEauthorblockA{\IEEEauthorrefmark{3}Georgia Institute of Technology, GA, USA}
    \IEEEauthorblockA{\IEEEauthorrefmark{4}Stony Brook University, NY, USA}
}

\maketitle

\begin{abstract}
High-fidelity direct numerical simulation of turbulent flows for most real-world applications remains an outstanding computational challenge.
Several machine learning approaches have recently been proposed to alleviate the computational cost even though they become unstable or unphysical for long time predictions.
We identify that the Fourier neural operator (FNO) based models combined with a partial differential equation (PDE) solver can accelerate fluid dynamic simulations and thus address computational expense of large-scale turbulence simulations.
We treat the FNO model on the same footing as a PDE solver and answer important questions about the volume and temporal resolution of data required  to build pre-trained models for turbulence.
We also discuss the pitfalls of purely data-driven approaches that need to be avoided by the machine learning models to become viable and competitive tools for long time simulations of turbulence.

\end{abstract}


\begin{IEEEkeywords}
Machine Learning, Turbulence, Fourier Neural Operator
\end{IEEEkeywords}

\section{Introduction}
Machine learning (ML) has improved efficiency of forecasting in weather and climate models (\cite{lam2023learning,pathak2022fourcastnet}). 
However, long time forecasts of ML-based emulators have been found to become unphysical or unstable (see \cite{lai2024machine} and references therein). 
Recently, it was shown in Ref. \cite{chattopadhyay2023long}) that the reason for this failure of ML-based emulators is spectral bias, where the smaller scales are not learned and only the large-scale dynamics are captured. 
In this paper, we propose a hybrid emulator for isotropic two-dimensional turbulence, which combines ML predictions with the governing partial differential equations (PDE) to march forward in time.  
This approach when generalized can be applied for long term predictions in climate models which are notoriously complex to solve due to the high computational cost of numerical PDE solvers. 
The key idea is to trade speed of the pure ML-emulator with the accuracy of the hybrid framework, i.e., the hybrid framework is not as fast as the pure ML-emulator but alternating between ML-emulator and PDE solver ensures that the results of the hybrid framework remain physical.       

While the idea of using neural networks for flow field reconstruction has existed for over two decades \cite{milano2002neural}, the application of machine learning in fluid dynamics witnessed a massive surge a decade ago \cite{Ling_Kurzawski_Templeton_2016}. 
Several works have since explored data-driven turbulence models, flow field reconstruction near walls, flow pattern recognition, \review{machine learning accelerated simulations}, and super-resolution (see \cite{bhushan2020machine,Fukami_Fukagata_Taira_2021,kochkov2021machine,vinuesa2021potential,obiols2020cfdnet,wang2022finding,lino2023current} and references therein).
Although, there exist several ML-based approaches for \review{steady-state Reynolds-averaged Navier-Stokes simulations} and the interpolation of flow fields in space and time, extrapolation problems have remained hard to tackle. 
This is particularly due to the nonlinear, multiscale, and chaotic nature of turbulence which remains a computational challenge for several important real-world applications of Navier-Stokes equations.
The large memory requirement and scarcity of data have so far prohibited learning a generalized solution operator of the Navier-Stokes equations.
Nevertheless, neural operators and embedding physical constraints in machine learning have shown promise in accelerating computational fluid dynamics.


In this paper, we evaluate the merits and discuss the pitfalls of a certain ML-based model for predicting spatiotemporal fluid dynamics.
We choose to study decaying turbulence in two dimensions (2D) as it is an important problem and can be extended to forced turbulence or three dimensions.
In Sec. \ref{sec:fno} we briefly introduce the ML model.
We adopt a purely data-driven approach and couple it with the numerical PDE solver to contain the growth of errors.
It is unfair to expect a purely data-driven approach trained on a limited amount of data to accurately predict long term dynamics of a chaotic system. 
Thus, in Sec. \ref{sec:dataset} we discuss the data set and in Sec. \ref{sec:dataanalysis} assess the time over which a data-driven solver can be expected to make accurate predictions by calculating the Lyapunov exponents and computing evolution statistics of global quantities.
The key question is how much data (samples and time refinement) is needed for an ML model to become a viable alternate to numerical solvers. 
Purely data-driven approaches are unlikely to replace the physical law-based numerical solvers \cite{succi2019big,viswanath2023neural} as they lack the knowledge to model all relevant physics of a complex multiscale flow. 
Nevertheless, one needs to discuss its advantages and limitations by putting it on the same footing as numerical PDE solvers. 
To that end, in Sec. \ref{sec:results} we identify an ML model, quantify the time refinement and number of samples (with different initial flow fields) required to train an ML model, and couple the ML model with PDE solver to make accurate long time predictions of the flow field at the physical conditions \review{and Reynolds number} relevant to atmospheric physics \cite{gao2018investigation}.

    
    



\section{Neural operators and related work} \label{sec:fno}

\newcommand{\mcirc}{\, \circ \,}

A neural operator is a neural network architecture that is designed to approximate a solution operator of resolution-independent PDEs.  
They learn mappings between infinite dimensional function spaces using a finite set of input-output data \cite{chen_1995,li2020fourier,deeponet_2021,JMLR:v24:21-1524,zhang2024emulator}.
For example, the Fourier neural operator (FNO) operates in the frequency domain by learning mappings between Fourier coefficients of high-dimensional data \cite{li2020fourier,JMLR:v24:21-1524}.
The deep network operator (DeepONet) \cite{deeponet_2021} utilizes two subnetworks to encode solution operators to several deterministic and stochastic differential equations.
The Laplace neural operator has been shown to work well with non-periodic functions \cite{cao2024laplace}.
The Markov neural operators are used to learn chaotic dynamics of non-ergodic dissipative systems \cite{li2021learning}.
The Clifford neural operators employ correlations between fields to learn multivector fields \cite{brandstetter2022clifford}.
The Riemann operator network learns solution operators to hyperbolic PDEs with discontinuous solutions \cite{peyvan2024riemannonets}.
Additionally, there are several variations of each neural operator with their own special characteristics.

In this work, we select the Fourier neural operator (FNO) as it is a promising tool for learning discretization-agnostic approximations to the solution operator of differential equations.
FNO truncates the high-frequency modes that are considered to be less important.
Informally, if $\mathcal{A}$ and $\mathcal{U}$ are two Banach spaces of functions defined, respectively, on bounded domains $\Omega_{d_{a}} \subset \mathbb{R}^{d_{a}}$ and $\Omega_{d_{u}} \subset \mathbb{R}^{d_{u}}$, and $\mathcal{G}$ is an operator that maps functions $a \in \mathcal{A}$ to functions $u \in \mathcal{U}$, i.e., $\mathcal{G}: \mathcal{A} \rightarrow \mathcal{U}$, a FNO parameterizes the operator $\mathcal{G}$ by $\mathcal{G}_{\theta}: \mathcal{A} \times \theta \rightarrow \mathcal{U}$, such that for some $\theta^{*} \in \mathbb{R}^{p}$, we have the approximation $\mathcal{G}_{\theta^{*}}  \ \approx \  \mathcal{G}$.  Thus $\mathcal{G}_{\theta^{*}}$ acts as a surrogate to the operator $\mathcal{G}$.   In the case of PDEs, the space $\mathcal{A}$ comprises ``input'' functions that the PDEs depend on, such as initial or boundary conditions, while the operator $\mathcal{G}$ maps these inputs to the solution $u \in \mathcal{U}$ satisfying the PDEs with the given inputs.  In the present study, the input space $\mathcal{A}$ comprises initial conditions.

\section{The Data Set} \label{sec:dataset}


\begin{figure}
    \centering
    \includegraphics[trim={0cm 0cm 0cm 0cm},clip,width=0.23\textwidth]{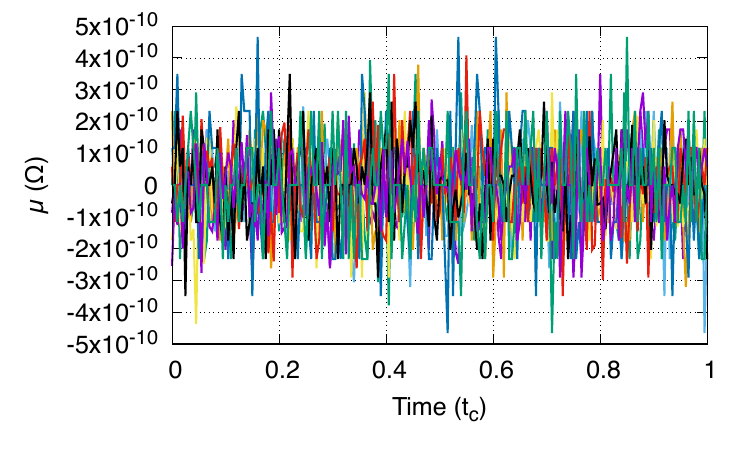}
    \includegraphics[trim={0cm 0cm 0cm 0cm},clip,width=0.24\textwidth]{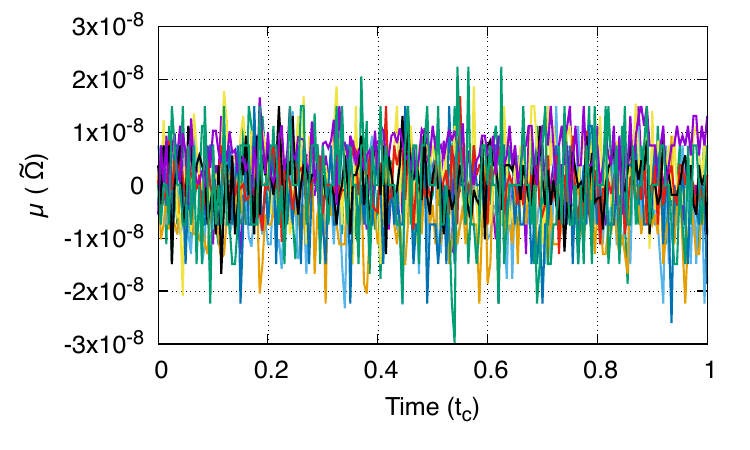}
    \\
    \includegraphics[trim={0cm 0cm 0cm 0cm},clip,width=0.23\textwidth]{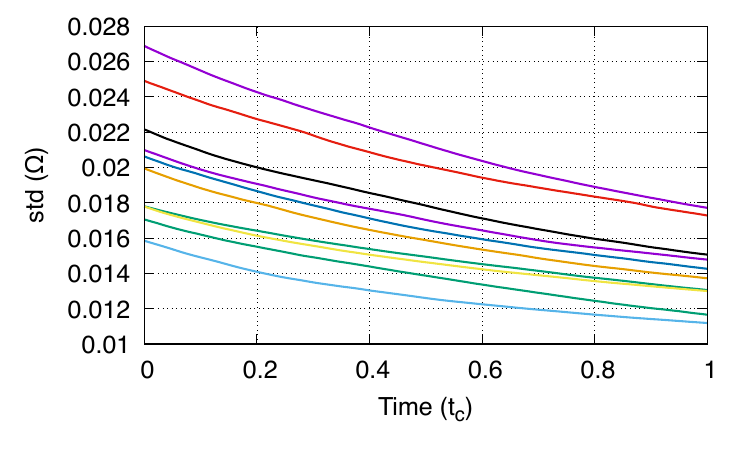}
    \includegraphics[trim={0cm 0cm 0cm 0cm},clip,width=0.23\textwidth]{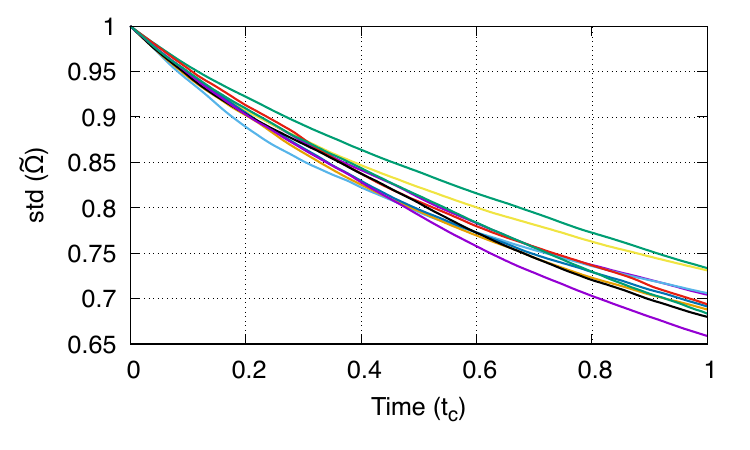}
    \\
    \includegraphics[trim={0cm 0cm 0cm 0cm},clip,width=0.23\textwidth]{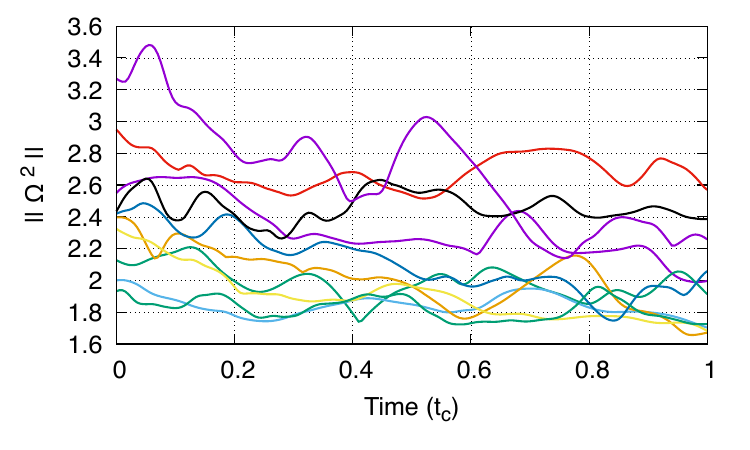}
    \includegraphics[trim={0cm 0cm 0cm 0cm},clip,width=0.23\textwidth]{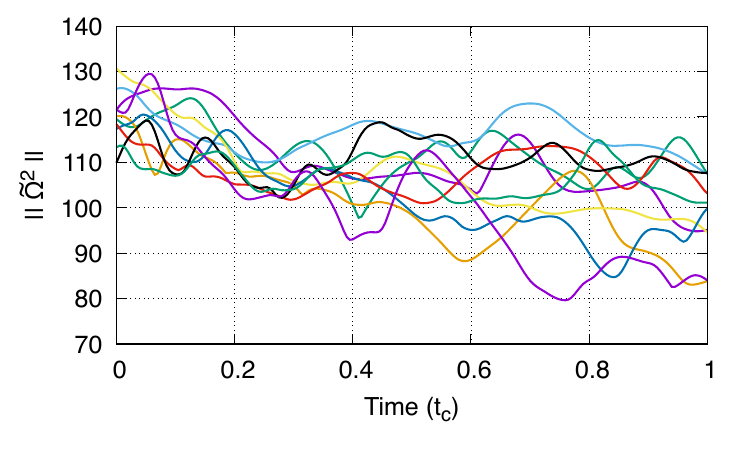}
    \caption{\label{fig:all_stats} Mean [top row], standard deviation [middle row], and Frobenius norm [bottom row] of raw vorticity ($\Omega$) [left column] and normalized vorticity ($\tilde \Omega$) [right column]. The normalization is with respect to mean and standard deviation from $t=0$. Each curve represents a different sample from the data set of 5000 samples.}
\end{figure}

The data set consists of vorticity and velocity fields from 5000 simulations of 2D decaying turbulence. 
These simulations differ from one another in that they are initialized with different uniformly distributed random numbers but the same transport coefficient. 
The initial condition generates several opposite vortices that diffuse as time progresses and convect around the periodic domain depending on their initial locations.
The Reynolds number is in the range of 7000-8000 for different samples depending on the initial condition.
We solve the Navier-Stokes equations for the velocity field using the entropic lattice Boltzmann method \cite{atif2017,atif2024lattice} on a grid of $256 \times 256$ points.
\review{The lattice Boltzmann method is an alternate methodology for computational fluid dynamics based on discrete space-time kinetic theory (see \cite{atif2022} for details). Each initial condition consumes 263 seconds on a single core of AMD EPYC 7402 CPU.}
The flow field is first allowed to evolve for $0.5\, t_c$ (where $t_c = L/U_0$ is the convection time and $U_0$ is the characteristic velocity) so that the initial sharp discontinuities vanish.
Thereafter, time is reset to zero and the velocity (${\bf u}$) and vorticity ($\omega_z$) are sampled from time $t=0$ to $t=t_c$ in steps of $0.005 \, t_c$.
Note that the vorticity field $\omega_z(x,y)$ is calculated as the curl of the velocity using $\omega_z(x,y) = \nabla \times {\bf u}(x,y)$, where $\nabla$ is the nabla operator and the velocity vector ${\bf u}$ has components $u_1, u_2$ for two-dimensional fluid flow.
Figure \ref{fig:ic_turb_sample} visualizes the two dimensional vorticity field of one such sample where several inversely rotating vortices are visible.
In general, in an incompressible viscous flow the vortices stretch (absent for two dimensions) and turn under the influence of each other's field and diffuse as the time advances.

\begin{figure}
\begin{tikzpicture}
    \node [] at (0, 0) { \includegraphics[trim={0cm 0cm 0cm 0cm},clip,width=0.4\textwidth]{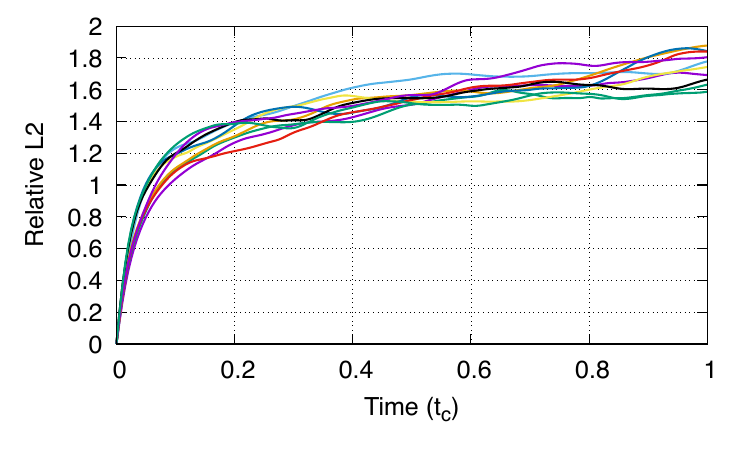} };
    \node [] at (1.8, -0.05) { \includegraphics[trim={0cm 0cm 0cm 0cm},clip,width=0.25\textwidth]{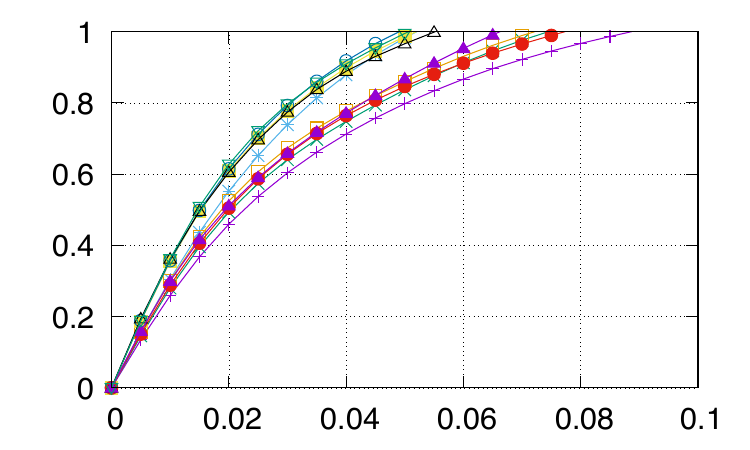} };
\end{tikzpicture}
\caption{\label{fig:normalizedl2} $L_2$ norm of difference of vorticity fields of ten samples with their respective initial values.}
\end{figure}

\begin{figure}
\begin{tikzpicture}
    \node [] at (0, 0) { \includegraphics[trim={0cm 0cm 0cm 0cm},clip,width=0.4\textwidth]{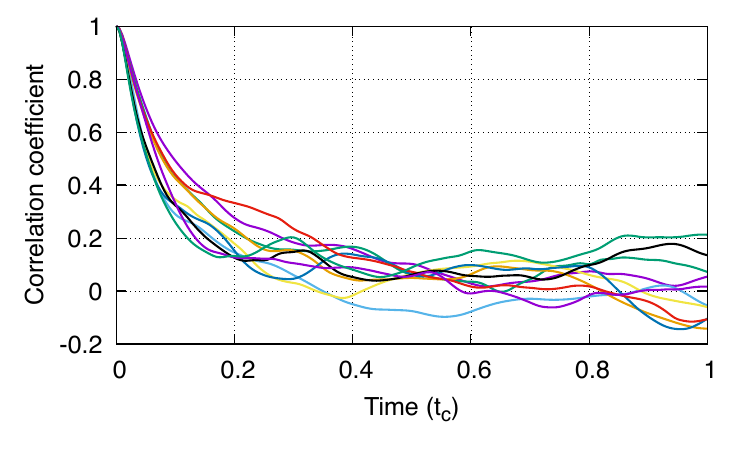} };
    \node [] at (1.8, 1.1) { \includegraphics[trim={0cm 0cm 0cm 0cm},clip,width=0.25\textwidth]{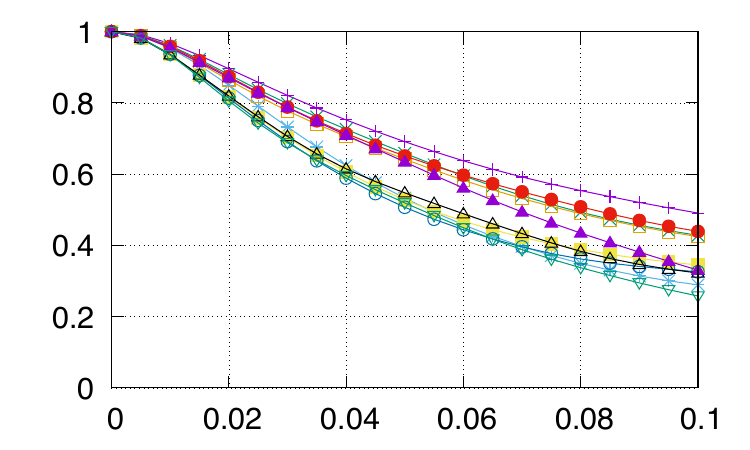} };
\end{tikzpicture}
\caption{\label{fig:normalizedcorr} Normalized projection of vorticity fields of the sample data sets on their respective initial values.}
\end{figure}



\begin{figure}
    \centering
    \includegraphics[trim={0cm 0cm 0cm 0cm},clip,width=0.3\textwidth]{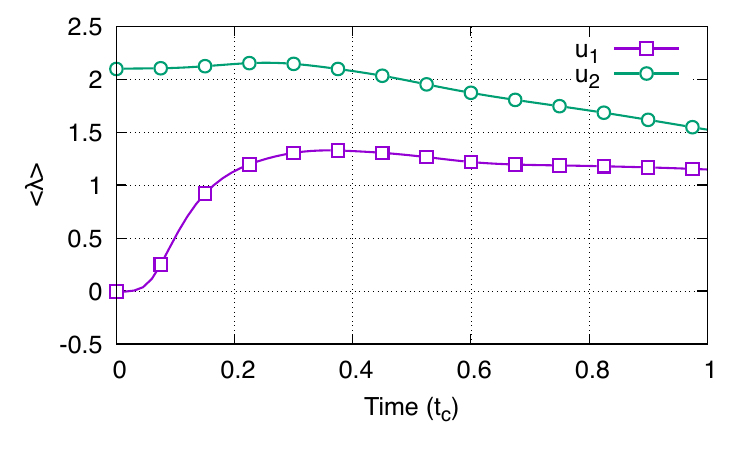}
    \caption{Lyapunov exponents calculated for the two components of the velocity vector.
    }
    \label{fig:lyapunov_time}
\end{figure}

\section{Spatiotemporal Characteristics of the Flow Field} \label{sec:dataanalysis}

In general, turbulent dynamics are chaotic. 
Chaotic systems are characterized by sensitivity to initial conditions, i.e., two nearby initial conditions separate exponentially even when the underlying dynamical equations are deterministic (\cite{strogatz2018nonlinear}, \cite{cvitanovic2005chaos}). 
For a given chaotic system, the Lyapunov time gives a time scale over which dynamics are predictable.
Data-driven methods need to be mindful of this time scale to ensure accuracy and confidence when extrapolating in time. 
In this section, we first analyze the data set of 2D turbulence to understand the spatiotemporal characteristic of the flow field by computing the evolutions of the correlation coefficient and relative $L_2$ errors.
The analysis ensures that there has been a meaningful evolution or separation from initial condition so that the prediction holds value, as, one pitfall is to make extremely short time predictions when the fields have evolved by such a tiny amount that even the initial condition would be an acceptable prediction.
This is followed by estimating the Lyapunov time scale.
 
The evolution of volume mean of vorticity, standard deviation of vorticity, and global enstrophy (defined as the sum of square of vorticity fluctuation over the domain) with time for a subset of the full dataset has been shown in Fig. \ref{fig:all_stats}. 
The left column describes the behaviour of a data set as is, while the right column consists of the normalized datasets where the vorticity at each point in a sample has been scaled with the vorticity field's mean and standard deviation at the initial instant. 
The mean remains constant at zero due to incompressibility (thus vorticity fluctuation is the same as vorticity), while the standard deviation decays with time as is expected of decaying 2D turbulence. 
The global enstrophy computed from normalized vorticity also decays as small scale structures are dissipated due to viscosity. 
In order to emphasize the turbulent nature of the flow and determine time scales associated with chaotic dynamics, we plot in Figs. \ref{fig:normalizedl2} and \ref{fig:normalizedcorr} two measures of time evolution for ten samples from the full dataset. 
Fig. 2 shows the separation between the vorticity fields for each dataset at time $t$ from its initial value at time $t=0$, scaled with their respective initial values. 
In Fig. 3, we plot the correlation coefficient of the vorticity field after time $t$ and the initial vorticity field for the data sets scaled with their respective initial values. 
As expected, this correlation coefficient between the vorticity fields decays with increasing time. 

For a dynamical system $\dot{\boldsymbol{x}} = \boldsymbol{f(x)}$, in practice, we select two nearby initial conditions with initial separation denoted by $\delta x_0 = ||\delta \boldsymbol{x}(t = 0)||$, with a suitable norm chosen as measure of distance in the state space. 
As these trajectories evolve in time, we track their separation $\delta x (t) = ||\delta \boldsymbol{x}(t)||$ as a function of time. 
Sensitivity to initial conditions dictates that nearby initial conditions separate exponentially and can be expressed mathematically as $\delta x (t) \approx (\delta x_0) e^{\lambda t}$ . 
In this work, we define Lyapunov time as $T_L = 1/\Lambda$, where $\Lambda$ is the maximum Lyapunov exponent defined as $\Lambda = \textrm{max} <\lambda>$, where
\begin{equation}
<\lambda> = \frac{\sum_i \lambda_i t_i}{\sum_i t_i},
\end{equation}
with  $\lambda = ({1}/{t}) \ln{({\delta x(t)}/{\delta x_0})}$.
To calculate $\Lambda$, we started with two initial conditions $A$ and $B$ such that $\delta x_0 = ||u^{(A)}_1(t=0) - u^{(B)}_1 (t=0)|| = 10^{-2}$, with $|| \cdot ||$ being the $L_2$ norm.
We then trace the evolution of $u^{(A)}_1,u^{(A)}_2,u^{(B)}_1,u^{(B)}_2$ and calculate the quantity $\lambda_i$ at each time-step $t_i$ for $u_1$ and $u_2$. 
Since the chaotic system has a finite maximum extent, the quantity $\lambda_i$ grows as trajectories separate exponentially under chaotic dynamics until the separation between them becomes the size of the chaotic attractor itself in the state space. 
The larger of the two $\Lambda$'s is $\approx 2.15$ while their average is $\approx 1.7$. 
Thus, a conservative estimate of the Lyapunov time is $T_L = \Lambda^{-1}  \approx 0.45$ convective time--units. 
This is consistent with Fig. \ref{fig:normalizedcorr} where vorticity correlation coefficients are seen to decay until $T_L$ after which the trajectories become independent.
We will therefore restrict the predictions from FNO to a time horizon smaller than $T_L$.


\section{Methods} \label{sec:methods}

In this section, we briefly introduce the methods for spatiotemporal dynamics of decaying turbulence.
The goal is to find the best methodology for extrapolating the flow field in time from a few inherently chronological snapshots.
The three methods that we consider in this paper are:
\begin{itemize}
    \item \textbf{2D FNO with temporal channels: } Here, the two dimensions of FNO model spatial features. This is augmented with channels (similar to the concept of RGB channels in a convolutional neural network for image data \cite{krizhevsky2012imagenet}) to include the time dimension \review{where the channels are chronologically ordered}. The number of input channels and output channels are parameters equal to number of input and output time snapshots respectively.
    \item \textbf{3D FNO:} In 3D FNO, two dimensions model the spatial features and one dimension models the temporal features. This does not differentiate between the spatial and temporal dimensions.
    \item \textbf{Hybrid FNO-PDE: } This method alternates between FNO and the PDE solver by using the output of FNO as input to PDE solver and vice versa. This could use either of the two FNO models listed above. It combines the speed of machine learning with the accurate physics of PDE solver. We leverage \verb|torchScript| to execute this workflow \cite{devito2022torchscript}.
\end{itemize}
In the above methods, we use PyTorch's neural operator library \cite{neuralop} for FNO and particle-resolved direct numerical simulation \cite{gao2018investigation} as the \verb|C++| PDE solver.

\section{Results} \label{sec:results}

In this section, we study the errors of different FNO models for the spatiotemporal dynamics of decaying 2D turbulence. 
It should be noted here that the traditional PDE solvers have well understood time-stepping thresholds rooted in their stability analyses and requirements to capture the smallest scales of turbulence.
However, in absence of an analogous thresholds for data-driven models one needs to rely on numerical experiments to decide the time resolution required for an accurate solution.
Our previous study has demonstrated that a minimum of ten snapshots with a time resolution of $0.05 \, t_c$ is necessary for making predictions at Reynolds number of 1000 \cite{atif2024towards}.
In this paper, we increase the Reynolds number by approximately $10$ times and refine the time resolution by a factor of $10$.
We will evaluate the models in terms of their number of parameters, training time, and error accumulated in long time roll-outs.
In all the following studies, unless otherwise noted, the errors are reported as an average of 500 samples (all with different initial conditions) that were not a part of the training data.

\begin{table*}[]
    \centering
    \begin{tabular}{c|c|c|c|c|c|c}
    Model                 & Width & Layers & Modes &  Time (hours) & Training Size & Parameters\\
    \hline\\

    2D FNO + Channels (10)&  40   &   4    &  32   & 2.41          & 1,000          & 6,995,922\\
    2D FNO + Channels (10)&  8    &   4    &  32   & 1.36          & 1,000          & 288,562 \\
    
    2D FNO + Channels (5) &  40   &   4    &  32   & 7.25          & 6,000          & 6,994,637\\
    2D FNO + Channels (5) &  8    &   4    &  32   &  4.07         & 6,000          & 287,277\\
  
    2D FNO + Channels (1) &  40   &   4    &  32   &  11.48        & 10,000        & 6,993,609  \\ 
    2D FNO + Channels (1) &  8    &   4    &  32   &   6.18        & 10,000        & 286,249    \\
    
    3D FNO                &  40   &   4    &  32   &  23.38        & 1,000          & 222,850,505 \\
    3D FNO                &  40   &   4    &  16   &  10.09        & 1,000          & 29,519,305 \\
    3D FNO                &  20   &   4    &  24   &  14.01        & 1,000          & 23,974,565 \\
    {3D FNO}       &  8    &   4    &  32   &  10.06        & 1,000          & 8,918,313 \\
    3D FNO                &  4    &   8    &  32   &  11.37        & 1,000          & 4,459,685 \\
    3D FNO                &  8    &   8    &  24   &  12.54        & 1,000          & 7,673,417 \\

    \end{tabular}
    \caption{\label{tab:params} Total number of parameters in the model and their training time on Nvidia A6000. }
\end{table*}

\subsection{2D FNO with temporal channels}

The 2D FNO model uses Fourier modes in the two spatial dimensions while stacking the time snapshots across channels with an inherent chronological ordering for temporal dimension. 
During training, the number of input channels is fixed to ten, but the number of output channels is varied from one to ten. 
The 2D FNO is used iteratively by using the outputs of the previous time as the input to ensure ten time snapshots are available as output when comparing errors.
In order to ensure fairness, the models have been trained on equal volume of data. 
Note that when using fewer output channels, more training samples are generated using the same data volume.


\begin{figure}
    \centering
    \includegraphics[trim={0cm 0cm 0cm 0cm},clip,width=0.35\textwidth]{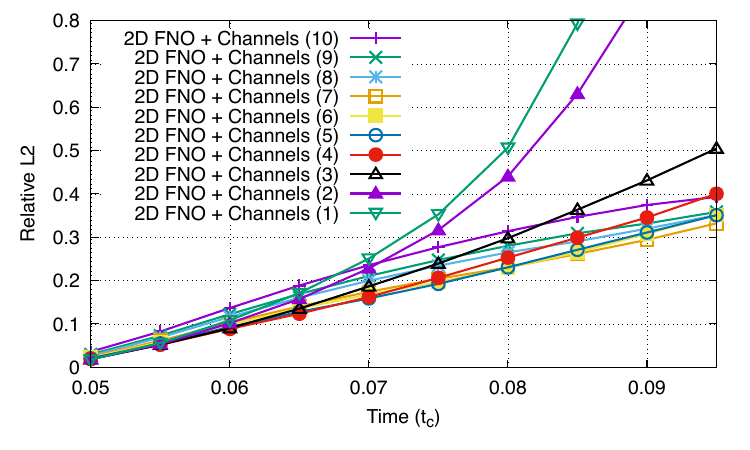}
    \includegraphics[trim={0cm 0cm 0cm 0cm},clip,width=0.35\textwidth]{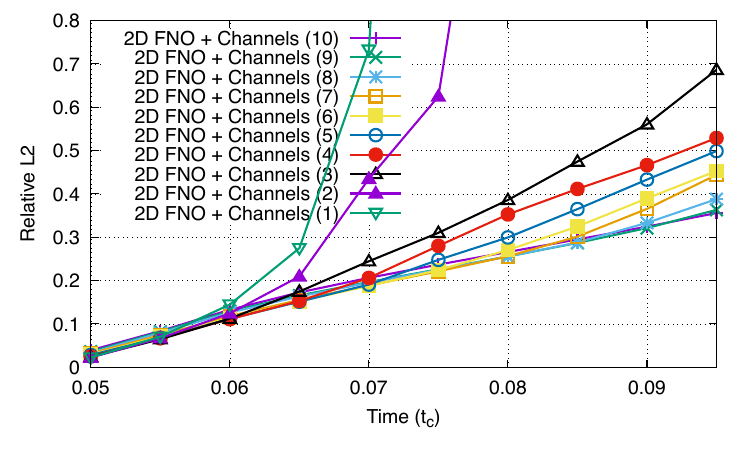}
    \caption{Different number of channels with width 8 (top) and width 40 (bottom) while the other hyperparameters are: layers (4), modes (32), scheduler gamma (0.5), scheduler step (100), and learning rate (0.001).}
    \label{fig:all_channels}
\end{figure}

\begin{figure}
    \centering
    \includegraphics[trim={0cm 0cm 0cm 0cm},clip,width=0.4\textwidth]{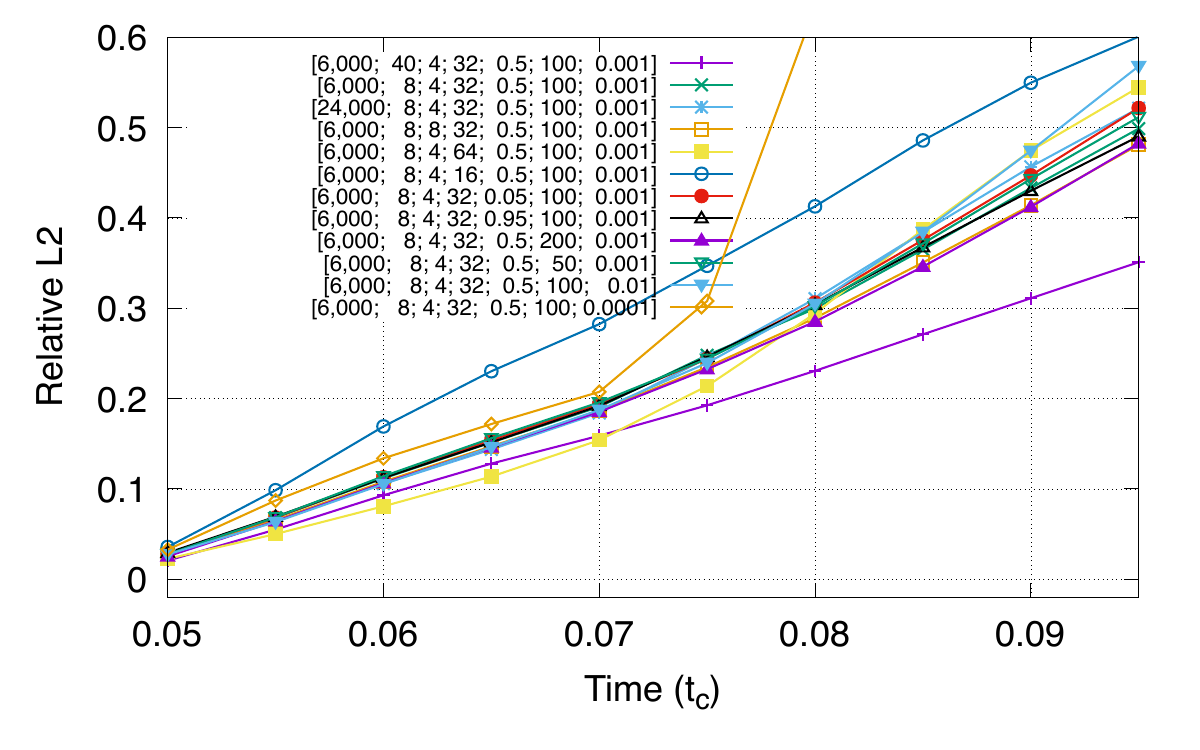}
    \includegraphics[trim={0cm 0cm 0cm 0cm},clip,width=0.4\textwidth]{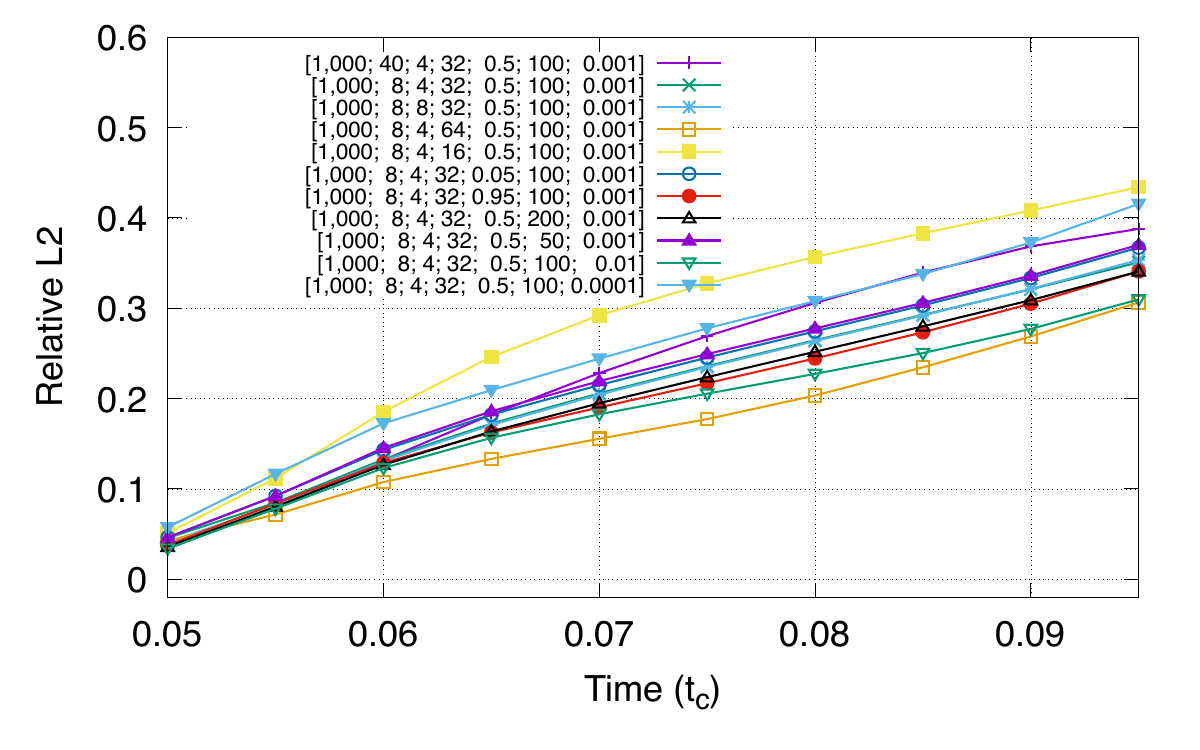}
    \caption{Hyperparameter tuning for Channels (5) and (10): samples, width, layers, modes, scheduler gamma, scheduler step, learning rate.}
    \label{fig:autoreg5_hyperparam_l2}
\end{figure}

We first plot the errors for different number of output channels in Fig. \ref{fig:all_channels} for two widths 8 and 40.
We notice that the errors are the largest for one output channel due to the ``compound error'' problem where prediction errors accumulate.
The errors with width 40 for the same number of output channels are in general higher which suggests overfitting. 
The errors for channels 5 and 10 are compared in Fig. \ref{fig:autoreg5_hyperparam_l2} for a wide range of hyperparameters from where it is seen that the errors are most sensitive to the number of Fourier modes.





\subsection{3D FNO}

\begin{figure}
    \centering
    \includegraphics[trim={0cm 0cm 0cm 0cm},clip,width=0.4\textwidth]{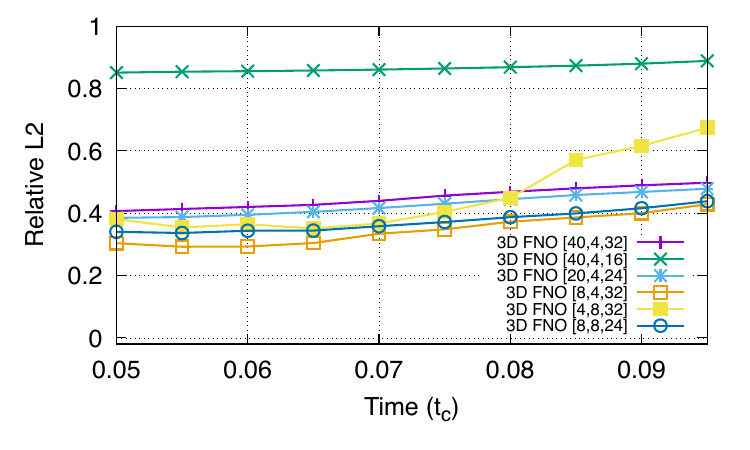}
    \caption{\label{fig:3dfno} Hyperparameter tuning of 3D FNO -- the three most important hyperparameters are the width, number of layers, and number of Fourier modes in each dimension.}
\end{figure}

The 3D FNO models accept Fourier modes in all dimensions (2 spatial, 1 temporal) with 10 time snapshots each as input and output. 
Figure \ref{fig:3dfno} compares the errors for several sets of hyperparameters.
It is seen that the errors are most sensitive to the number of Fourier modes.
It is also noticed that reducing the width improves the accuracy due to fewer parameters in the model (see Table \ref{tab:params}) which avoids the over-fitting problem.
It is interesting to note from Figure \ref{fig:3dfno} that the errors in 3D FNO models show weak dependence on time, in that they begin with large values and increase marginally as time progresses.
From Table \ref{tab:params} it can also be noted that the training time of 3D FNO is larger than 2D FNO with channels.
Therefore, 2D FNO with channels comes across as a better methodology for 2D turbulence due to its lower training time (hence the computational cost) and significantly lower errors for initial time steps.

\subsection{Hybrid FNO-PDE for long-term temporal stability}
\begin{figure*}[t]
    \centering
    \includegraphics[width=0.9\textwidth]{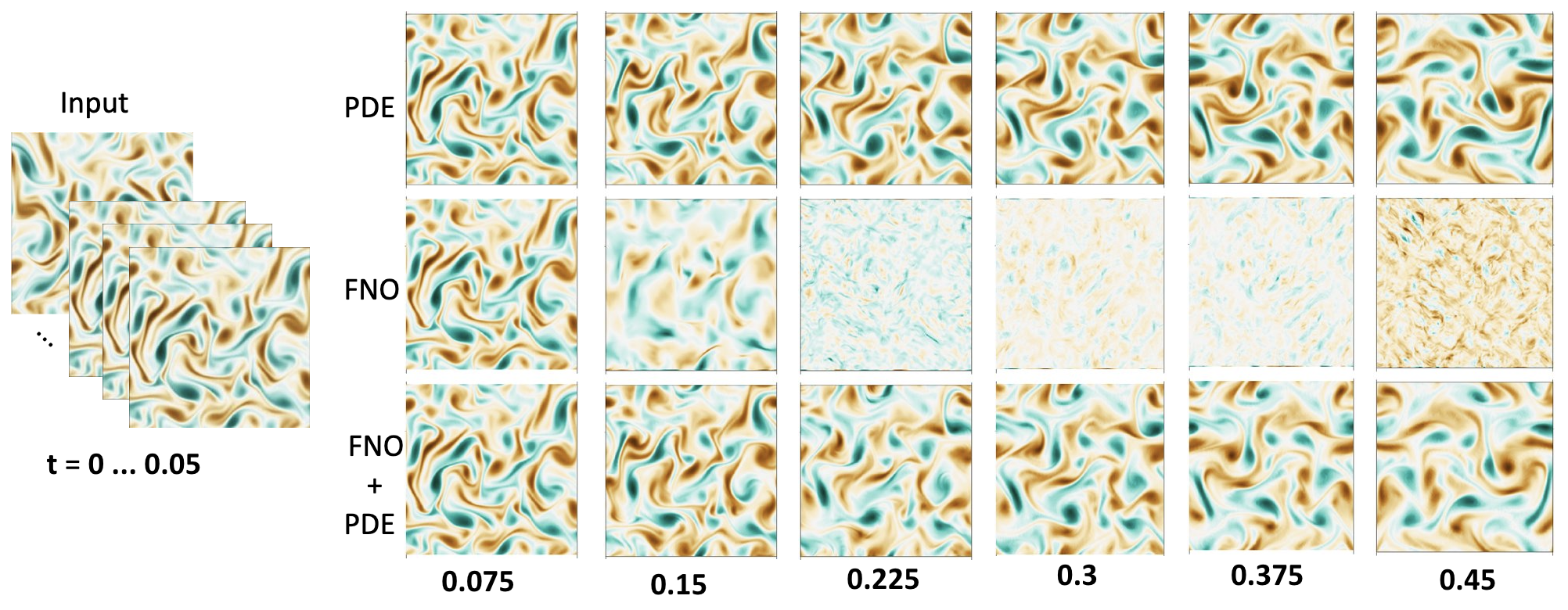}
    \includegraphics[trim={0cm 0cm 0cm 0cm},clip,width=0.32\textwidth]{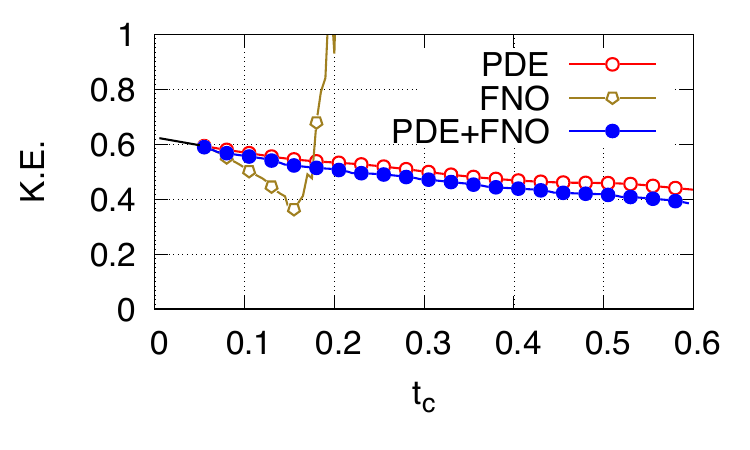}
    \includegraphics[trim={0cm 0cm 0cm 0cm},clip,width=0.32\textwidth]{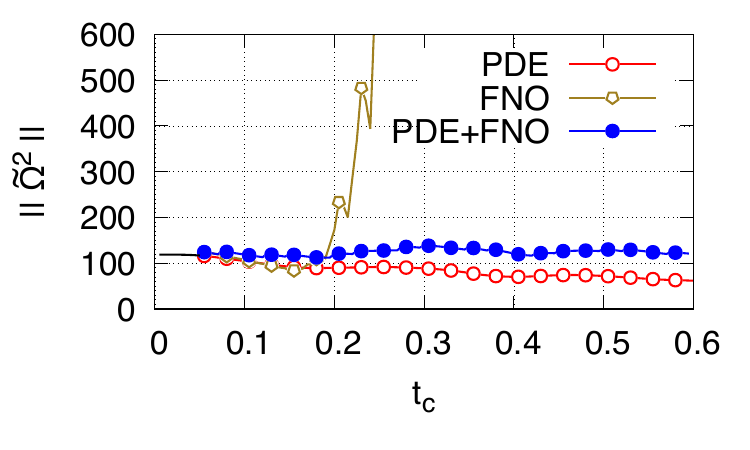}
    \includegraphics[trim={0cm 0cm 0cm 0cm},clip,width=0.32\textwidth]{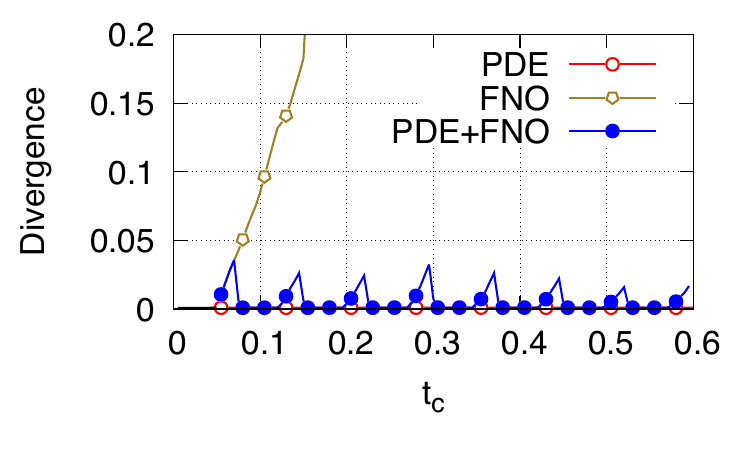}
    \caption{Visualization of long time predictions of vorticity fields from three models [top], and global statistics of kinetic energy (K.E.), global enstrophy ($||\tilde \Omega^2||$), and divergence ($\nabla \cdot \bm{u}$) [bottom]. Times are in units of convection time $t_c$. }
    \label{fig:ic_turb_sample}
\end{figure*}

\begin{figure}
    \centering
    \includegraphics[trim={0cm 0cm 0cm 0cm},clip,width=0.3\textwidth]{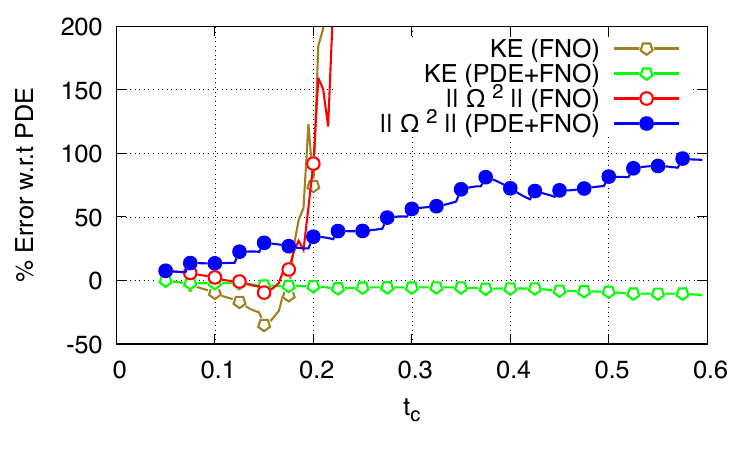}
    \caption{Percentage errors in kinetic energy (K.E.) and enstrophy ($||\Omega^2||$) for long time predictions.}
    \label{fig:ke_ens_longtime}
\end{figure}

As mentioned earlier, the aim of this paper is to assess the feasibility of FNO based models to augment numerical PDE solvers.
Thus in this section we investigate the error accumulation of a hybrid FNO+PDE methodology and compare it with pure FNO and PDE solver.
In the hybrid scheme, a single solver (numerical or ML-based ones) is invoked during a certain time window, and its output is used as the input of the other solver to make predictions for the next time window.
These two solvers are used iteratively until the whole time period is solved.
We choose the 2D FNO with 10 input channels (time $t = 0$ to $0.045 t_c$) and 5 output channels ($t = 0.05 t_c$ to $0.07 t_c$) with the best set of hyperparameters for this comparison.
The model is trained on velocity fields but we compute vorticity from the predicted velocity fields to ease visualization.
This model thus acts as a pre-trained ML model for decaying 2D turbulence.
Figure \ref{fig:ic_turb_sample} visualizes vorticity fields from the three methodologies -- PDE, 2D FNO with channels, and hybrid FNO-PDE.
Figure \ref{fig:ic_turb_sample} also plots the global values of kinetic energy, enstrophy, and divergence under different schemes for one sample.
It is observed that the predictions from FNO are not divergence free (as the incompressibility of velocity fields was not incorporated in the loss function while training), but the PDE solver drives the fields toward divergence-free condition.
Figure \ref{fig:ke_ens_longtime} plots the errors of kinetic energy and enstrophy for long term predictions and shows that errors from pure FNO get out of bound quickly while those from hybrid FNO + PDE remain stable.
It is interesting to note that errors in kinetic energy remain smaller than $10\%$ while the errors in enstrophy grow.
This can be attributed to the fact that enstrophy is calculated from the gradient of velocity field while the model lacks any explicit mechanism to learn gradients.
\review{This could be addressed by incorporating governing equations in the loss functions or applying a relevant filter and will be investigated in future works.}

\section{Discussion and Outlook} \label{sec:disc}

The numerical experiments in this paper lead to a viable hybrid methodology for accelerating direct numerical simulation by using FNO.
We have demonstrated that the FNO when trained on a sufficiently large data set leads to a pre-trained model which when coupled with the numerical PDE solver can be used to make stable long-term predictions.
The 3D FNO models were found to have a longer training time and yield larger errors for decaying 2D turbulence because the periodic spatial dimension cannot be treated at the same footing as non-periodic temporal dimensions.
Thus, we conclude that stacking the temporal dimension across channels leads to smaller errors and requires a shorter training time.

It should also be noted that the FNO model generalizes well as it was trained on data set generated using lattice Boltzmann but coupled with finite difference based Navier-Stokes solver.
This was possible as the physical conditions for both the numerical methods were identical by design, however, a word of caution is in order as the
generalizability of pre-trained machine learning models has been discussed recently in the context of ``foundational" models.
The FNO models discussed in this work have been trained on the data of decaying 2D turbulence for a specific range of Reynolds number.
In order to generalize it further as a solution operator for Navier-Stokes equation one needs to embed more physics in the training.
Foundational models to be trained on sufficiently diverse data set and should at the minimum replicate canonical test cases of fluid dynamics.
Their training also needs to incorporate the knowledge of governing equations as explored in many physics-informed machine learning works \cite{karniadakis2021physics,li2024physics}.
\review{An extension of the present framework to 3D should be straightforward with 3D FNO for spatial and channels for temporal dimensions}.

High-fidelity direct numerical simulation of turbulent flows for most real-world applications remains an outstanding challenge due to its computational cost.
Although, the advantage of machine learning inference manifests in the form of reduced time-to-solution, there exist several costs that are not frequently discussed such as -- (i) the computational costs of training a model, (ii) procuring a GPU, and (iii) energy cost, some of which are amortized over several inference steps.
In our hybrid PDE-FNO model presented here, the PDE solver takes 20 secs on AMD EPYC 7413 24-Core Processor for evolving flow field over $0.025 \, t_c$.
The machine learning component takes 0.1 secs for host to device data transfer and 0.3 secs for FNO inference on Nvidia A6000 GPU.
However, the degree of optimization of the software libraries as well as scalability of the methodologies vary.
Most high performance computing (HPC) codes are written in C++ while machine learning frameworks are Python based. 
While coupling the two one also needs to account for the cost of transforming the data from C++ dynamic memory arrays into libTorch tensors.
Several components discussed above often require optimization for real-world HPC applications which incur costs in terms of software redevelopment efforts.
Thus, a fair comparison of the computational cost of different methodologies accounting for a wide range of metrics needs more research.

\section{Acknowledgements}
The authors gratefully acknowledge financial support from the Laboratory Directed Research and Development program at Brookhaven National Laboratory, which is sponsored by the US Department of Energy, Office of Science, under  Contract Number DE-SC0012704.
This research used resources of the National Energy Research Scientific Computing Center (NERSC), a Department of Energy Office of Science User Facility using NERSC award ASCR-ERCAP0027399.

\bibliographystyle{IEEEtran}
\bibliography{main}

\end{document}